# Thermo-mechanical Characterization of 2D hexagonal Boron Phosphide (h-BP)

Zahabul Islam[1,*]

[1*]Mechanical and Manufacturing Engineering Program, School of Engineering, Bowling Green State University, Bowling Green, OH, 43403, USA

## Abstract

This study explores the thermo-mechanical properties of a two-dimensional (2D) monolayer hexagonal boron phosphide (h-BP). h-BP is predicted to possess a moderate band gap, high thermal stability, and excellent carrier mobility, making it suitable for advanced electronic, sensing, and energy applications. A classical molecular dynamics (MD) potential for h-BP was developed using density functional theory (DFT) calculations. The derived parameters were implemented in MD simulations to evaluate mechanical behavior under tensile loading along both zigzag and armchair directions at varying temperatures (300 K to 900 K). The results reveal significant anisotropy in mechanical performance, with higher tensile strength and elastic modulus in the zigzag direction across all temperatures. Increasing temperature reduces both tensile strength and stiffness due to thermal softening and increased atomic vibrations. The influence of structural defects was also investigated, revealing that Stone–Wales and vacancy defects reduce the tensile strength and failure strain of h-BP, with the two-atom vacancy producing the most pronounced mechanical degradation due to localized stress concentration and premature crack initiation. These findings provide a foundation for future research on the mechanical stability of h-BP in extreme environments.

* Corresponding author.

E-mail address: mdzisla@bgsu.edu

## 1. Introduction

Two-dimensional (2D) materials, such as graphene, molybdenum disulfide ($MoS_2$), and tungsten disulfide ($WS_2$), have attracted extensive attention due to their exceptional electronic, optical, and mechanical properties [1-4]. These materials typically consist of atomically thin layers with strong in-plane bonds and weak van der Waals interactions between layers, leading to unique characteristics like high electrical conductivity, flexibility, and mechanical strength [5]. For instance, graphene is renowned for its remarkable conductivity and tensile strength, while $MoS_2$ and $WS_2$ exhibit semiconducting properties with potential for electronic and optoelectronic devices.

Among the emerging 2D materials, hexagonal boron phosphide (h-BP) has gained interest due to its graphene-like honeycomb structure and favorable mechanical and electronic properties [1, 6, 7]. Although h-BP has not yet been synthesized experimentally, theoretical studies predict that it possesses high thermal stability, a moderate direct band gap, and excellent carrier mobility [6, 8]. These characteristics suggest that h-BP could play a vital role in next-generation electronic devices, sensors, and energy applications [9-13]. Additionally, strong B-P bonds in h-BP and robust lattice structure make it a promising candidate for use in harsh environments.

Investigating the mechanical properties of 2D materials like h-BP is crucial for their practical application. Properties such as elastic modulus, tensile strength, and fracture toughness influence the reliability and durability of materials under mechanical stress. Understanding how h-BP responds to different temperatures and mechanical loads is especially important for device performance and long-term stability[14, 15].

Molecular dynamics (MD) simulation is widely used in materials modeling to predict various physical properties, including mechanical behavior, thermal stability, and electronic structure, by simulating atomic interactions under different conditions. It also plays a crucial role in understanding complex phenomena such as radiation damage, phase transformations, and defect evolution in advanced materials [16-29]. Although there is growing interest in h-BP, the development of a classical MD simulation potential is essential for detailed atomic-level studies. This limitation has hindered detailed studies on its behavior at the atomic level under varying conditions. In the present study, this gap is addressed by developing a forcefield for h-BP based on density functional theory (DFT) calculations, and the derived parameters for bond, angle, and dihedral interactions were then implemented in MD simulations using LAMMPS package [30]. This study not only provides the first classical MD potential for h-BP but also offers valuable insights into its mechanical behavior, paving the way for future research and applications of this h-BP 2D material.

## 2. Computational Details

### 2.1 Forcefield Developments

The forcefield parameters for hexagonal boron phosphide (h-BP) were derived using first-principles density functional theory (DFT) calculations using Dmol3 package [31-33]. The spin-polarized plane-wave method along with the ultrasoft pseudopotential approximation to simulate the interactions between ions and electrons were implemented. The exchange-correlation energy was treated using the Generalized Gradient Approximation (GGA) with the Perdew-Burke-Ernzerhof (PBE) functional [31-33]. The plane-wave basis set was configured with a cutoff energy of 400 eV. Geometry optimizations were carried out using the BFGS minimization algorithm, with convergence criteria set to a total energy tolerance of $10^{-5}$ eV per atom and a maximum force residue of 0.03 eV/Å. The self-consistent calculations achieved convergence within a tolerance of $10^{-6}$ eV per atom. The electronic configurations for boron and phosphorus atoms were defined as $2s^2 2p^1$ and $3s^2 3p^3$, respectively. The DFT calculations were performed using a molecular representation of monolayer h-BP designed to reproduce the local B–P bonding environment. The optimized molecular structure was systematically perturbed through B–P bond stretching, B–P–B and P–B–P angle bending, and B–P–B–P and P–B–P–B torsional deformation. The corresponding DFT energy profiles were fitted to Morse bond, harmonic angle, and harmonic dihedral potential functions, respectively. The resulting equilibrium bond distance, equilibrium bond angles, and force constants were subsequently implemented in the classical molecular dynamics force field. Parameters extracted from DFT calculations served as the foundation for developing a reliable classical forcefield used in molecular dynamics simulations to evaluate the mechanical properties of h-BP as shown in Table 1.

**Table 1.** Optimized Forcefield parameters obtained from DFT calculation

| **Morse Potentials** | **$D_0$ (eV)** | **α (1/Å)** | **$r_o$ (Å)** |
|---|---|---|---|
| | 5.0162 | 1.3657 | 1.8673 |

| **Angle type Harmonic** | $k_\theta$ **(eV)** | **$\theta_o$ (degree)** |
|---|---|---|
| B-P-B | 2.0673 | 124.1410 |
| P-B-P | 2.3820 | 127.722 |

| **Dihedral Harmonic** | $k_\varphi$ **(eV)** | **Phase (d)** |
|---|---|---|
| B-P-B-P<br>P-B-P-B | 0.4023 | -1 |

**Table 2.** LJ 12-6 Parameters for non-bonded interactions [34]

| **LJ 12-6** | **ε (eV)** | **σ (Å)** |
|---|---|---|
| B-B | 0.00781 | 4.0830 |
| B-P | 0.01016 | 4.1148 |

| P-P | 0.01322 | 4.1470 |
|---|---|---|

### 2.2 Classical MD simulations

The MD simulations were conducted using LAMMPS package [30]. The simulation time step was set as 1 fs and a pressure of 1 bar. The forcefields parameters were used from quantum calculation and the following interactions were considered during simulation [30, 34]:

$$E_{Total} = E_{bonded} + E_{non-bonded} \quad (1)$$

$$E_{Total} = E_{bond} + E_{angle} + E_{dihedral} + E_{vdW} \quad (2)$$

$$E_{bond} = D_o[e^{-2\alpha(r-r_{o})} - 2e^{-2\alpha(r-r_{o})}] \quad (3)$$

$$E_{angle} = k_\theta(\theta - \theta_o)^2 \quad (4)$$

$$E_{dihedral} = k_\varphi[1 + d\cos n\varphi] \quad (5)$$

$$E_{vdW} = 4\epsilon[\left(\frac{\sigma}{r}\right)^{12} - (\frac{\sigma}{r})^6] \quad (6)$$

Detailed description of the parameters can be found elsewhere [30, 34]. Further details of the molecular dynamics (MD) simulations are provided in reference [35]. The dimensions of the simulation cell were set to approximately 60 Å × 60 Å. A strain rate of $10^{-9}$ $s^{-1}$ was selected based on prior studies investigating the strain rate effect on mechanical properties [35, 36]. Prior to the tensile loading simulation cells were optimized using CG method and equilibrated under NPT dynamics for 100ps. This study examines the effect of different temperatures 300 K, 500 K, 700 K, and 900 K on mechanical behavior.

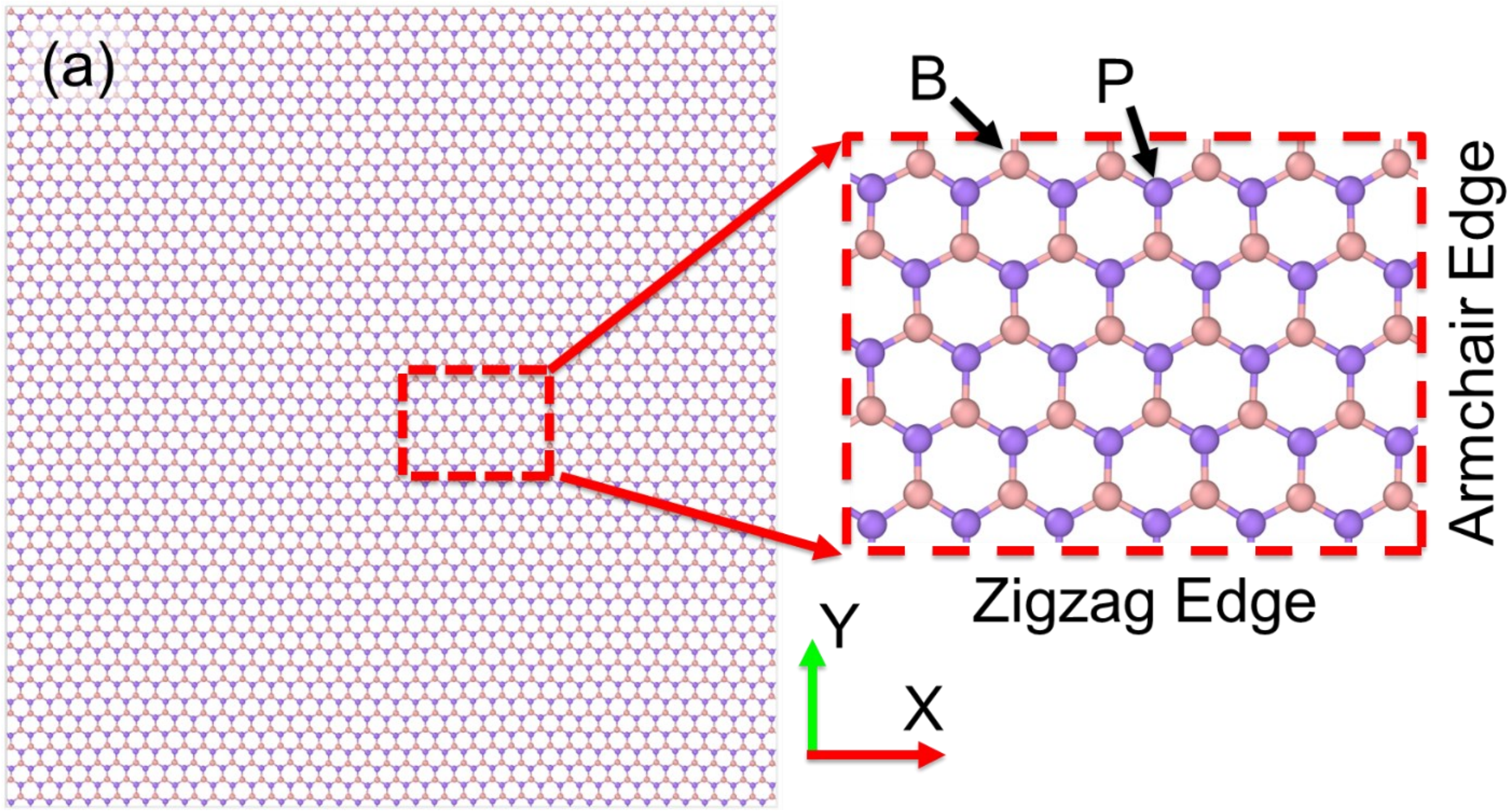

**Figure. 1**. (a) Computer model represents hexagonal monolayer h-BP.

## Results and Discussion

### 3.1 Tensile Loading in the Zigzag Direction

The stress-strain behavior of the monolayer hexagonal boron phosphide (h-BP) under tensile loading along the zigzag direction was studied at four different temperatures (300 K, 500 K, 700 K, and 900 K), as illustrated in Fig. 2. The stress-strain curves exhibit distinct trends that reflect the influence of temperature on the mechanical properties of h-BP. At 300 K, the monolayer demonstrates the highest tensile strength, reaching approximately 65 GPa at a strain of 26.28%. The material shows a characteristic linear elastic response followed by a sharp failure point at room temperature. The observed behavior suggests that the strong in-plane B-P bonds provide high mechanical stability under tensile loading. As the temperature rises to 500 K and 700 K, the peak stress decreases slightly, with values of around 64 GPa and 61 GPa, respectively. Similarly, the strain at failure reduces slightly, indicating that elevated temperatures weaken the interatomic bonds due to increased atomic vibrations, which promote earlier bond rupture. At 900 K, the stress-strain curve shows a more pronounced reduction in both peak stress (60 GPa) and strain-to-failure at 23.2%. The material becomes more susceptible to deformation and failure due to thermal softening, as increased atomic mobility weakens the lattice structure. Overall, the results confirm that the tensile strength and strain-to-failure of the h-BP monolayer decrease as temperature increases. This is consistent with the thermal degradation of mechanical properties observed in other 2D materials. These findings are critical for understanding the operational limits of h-BP in high-temperature environments and for designing applications that require both mechanical strength and thermal stability.

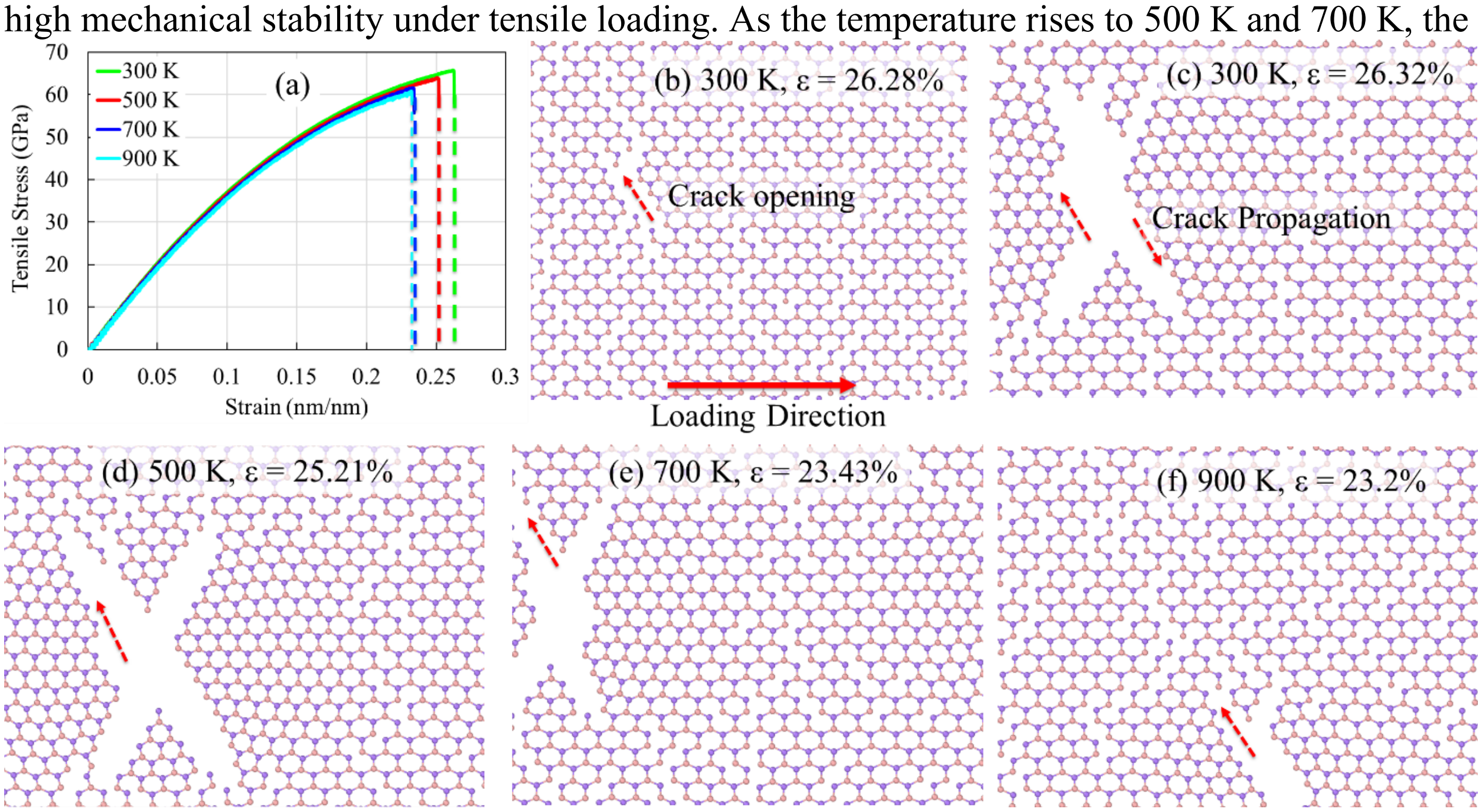

**Figure. 2.** Tensile response of h-BP at different temperatures along the (a) Zigzag direction loading, (b) at 300K, crack opening or initiation at 26.28% strain, crack propagation at: (c) 300K, and (d) 500K, (e) 700K, and (f) 900K.

### 3.2 Tensile Loading in the Armchair Direction

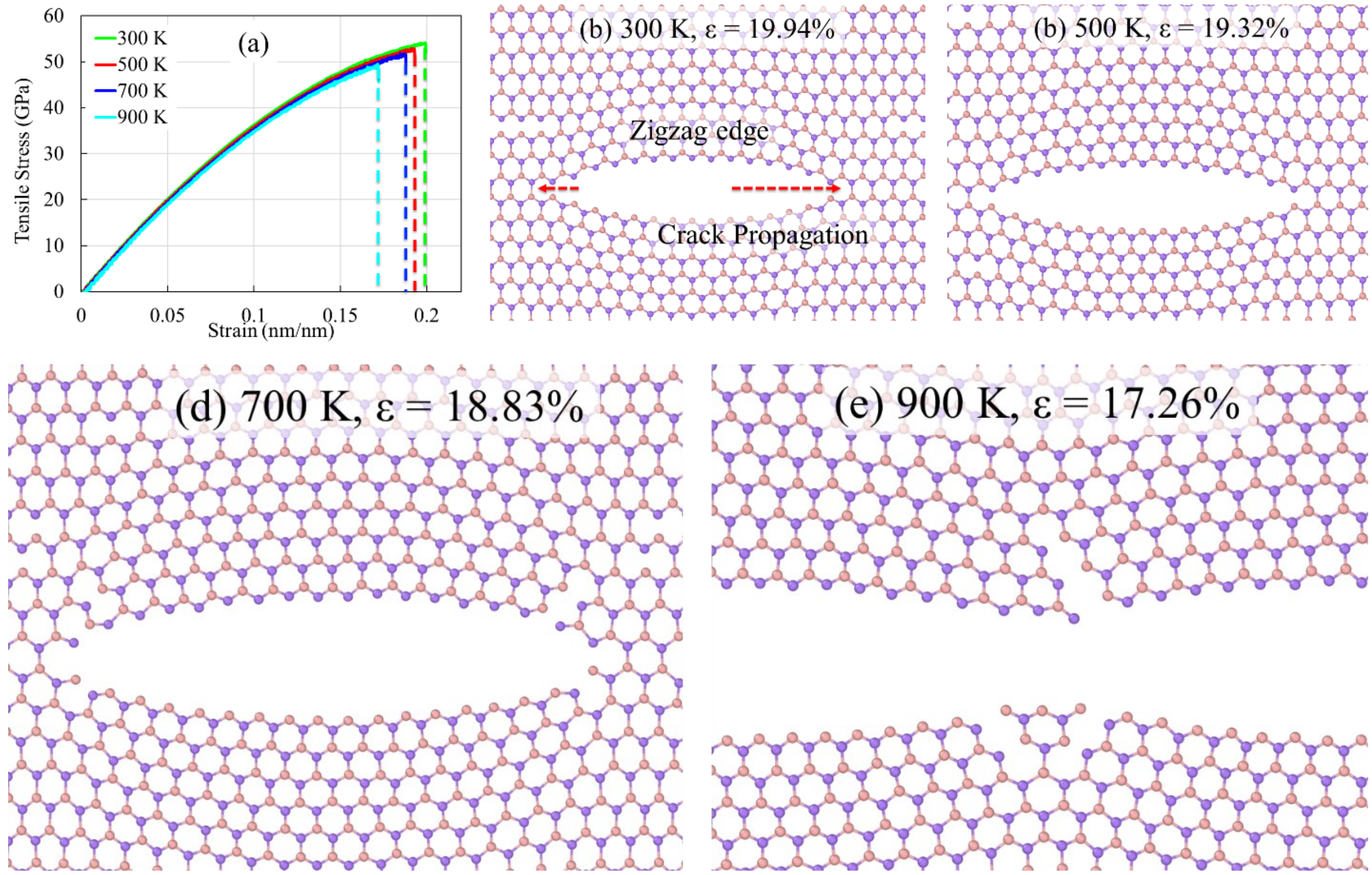

**Figure. 3.** Tensile response of h-BP at different temperatures along the (a) Armchair direction loading, (b) at 300K, crack opening or initiation, crack propagation at: (c) 300K, and (d) 500K, (e) 700K, and (f) 900K.

The tensile behavior of the monolayer hexagonal boron phosphide (h-BP) under armchair directional loading was studied at four different temperatures: 300 K, 500 K, 700 K, and 900 K, as shown in Fig. 3. The stress-strain curves exhibit a clear temperature-dependent reduction in mechanical performance. At 300 K, the stress-strain curve shows the highest tensile strength, with a peak stress of approximately 54.1 GPa at a strain of 19.94%. The material exhibits a predominantly elastic response followed by a sudden fracture, suggesting a brittle failure mode under room temperature conditions. At 500 K, the peak stress slightly decreases to around 52.9 GPa, while the strain at failure is also slightly reduced. This trend suggests that thermal agitation begins to affect the integrity of the B-P bonds, resulting in marginally weaker mechanical performance compared to room temperature. The behavior becomes more pronounced at 700 K, where the peak stress further drops to approximately 51.7 GPa. The corresponding strain at failure is reduced, indicating that thermal softening significantly impacts the mechanical stability at this temperature. At 900 K, the stress-strain curve shows a marked decrease in both peak stress (49.5 GPa) and strain-to-failure 17.26%. Comparing the mechanical response in the armchair direction to the zigzag direction, it is evident that tensile strength is slightly lower in the armchair direction across all temperatures. This anisotropy arises from the differences in atomic bonding and structural alignment in the two directions. Tensile properties of h-BP is found to be lower than 2D monolayer graphene [37] in both armchair and zigzag directions. Overall, these findings provide valuable insights into the thermal and mechanical behavior of h-BP, particularly under high-temperature tensile loading conditions in different crystallographic orientations.

### 3.3 Effect of Temperature on Tensile Strength and Elastic Modulus

#### 3.3.1. Temperature Dependence of Tensile Strengths

The effect of temperature on the tensile strength of hexagonal boron phosphide (h-BP) in both zigzag and armchair directions is shown in Fig. 4a. The data reveal a linear decrease in tensile strength with increasing temperature for both orientations. The tensile strength in the zigzag direction starts at approximately 65.7 GPa at 300 K and declines to around 60.8 GPa at 900 K. The linear relationship between temperature and tensile strength is described by the equation:

$$\sigma_{zigzag} = -0.0085 \times T + 68.2 \quad (7)$$

$$\sigma_{armchair} = -0.0075 \times T + 56.58 \quad (8)$$

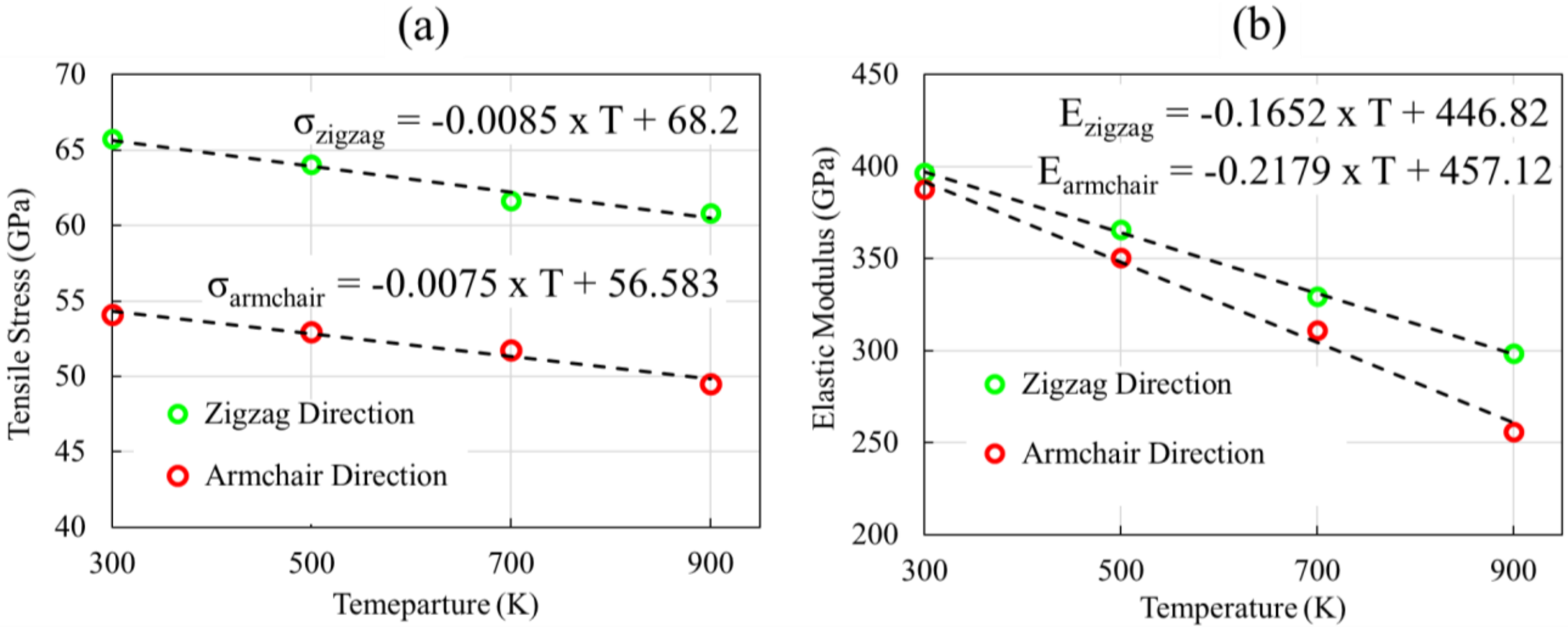

**Figure. 4.** Effect of Temperature on mechanical properties of h-BP: (a) Tensile strength, and (b) Elastic Modulus.

The eqn (7) indicates a temperature-dependent reduction of 0.0085 GPa per change in temperature (in K), highlighting the sensitivity of zigzag-direction mechanical properties to thermal effects. In the armchair direction, the tensile strength begins at 54.1 GPa at 300 K and decreases to about 49.5 GPa at 900 K. The relationship follows the eqn. (8). The slope of this line indicates a slightly lower reduction rate of 0.0075 GPa per Kelvin compared to the zigzag direction. These results demonstrate anisotropy in the mechanical behavior of h-BP, with the zigzag direction consistently exhibiting higher tensile strength than the armchair direction across all temperatures. This difference can be attributed to the inherent atomic arrangement and bonding in the two directions. The higher strength in the zigzag direction suggests that the B-P bonds are more effectively aligned to resist tensile deformation in this orientation. Both directions show a significant decrease in strength at elevated temperatures, which is consistent with the weakening of interatomic bonds due to increased thermal vibrations. This behavior aligns with the general mechanical response of other 2D materials under thermal loading. Understanding this temperature-dependent anisotropy is crucial for designing h-BP-based materials for applications where thermal stability and mechanical strength are critical.

### 3.3.2. Temperature Dependence of Elastic Modulus

The variation of the elastic modulus with temperature for h-BP in the zigzag and armchair directions is shown in Fig. 4b. The plot highlights a linear decline in elastic modulus as temperature increases, with distinct differences between the two crystallographic orientations. In the zigzag direction, the elastic modulus starts at approximately 396.8 GPa at 300 K and decreases to around 298.7 GPa at 900 K. The relationship between temperature and modulus is given by:

$$E_{zigzag} = -0.1652 \times T + 446.82 \quad (9)$$

$$E_{armchair} = -0.2179 \times T + 457.12 \quad (10)$$

The eqn. (9) indicates a reduction of 0.1652 GPa per change in temperature (K), reflecting the loss of stiffness at elevated temperatures. In the armchair direction, the elastic modulus begins at 387.9 GPa at 300 K and decreases to about 255.8 GPa at 900 K. The temperature-dependent behavior is described by eqn. (10) indicates modulus reduction rate is 0.2179 GPa per change in tempearture (K), which is higher than that observed in the zigzag direction. This suggests that the armchair direction experiences greater degradation in stiffness under thermal loading. Overall, the elastic modulus in both directions decreases significantly with temperature, indicating thermal softening of the material. The anisotropy in mechanical properties is evident, with the zigzag direction maintaining higher stiffness than the armchair direction throughout the temperature range. This behavior can be attributed to the structural arrangement and bonding characteristics of the B-P atoms along different orientations. Applications requiring high stiffness and thermal stability would benefit from leveraging the superior properties of the zigzag direction.

### 3.3.3. Effect of Defect on Mechanical Properties

The tensile stress–strain responses of pristine h-BP and h-BP containing a Stone–Wales (SW) defect, a single-atom vacancy, and a two-atom vacancy are shown in Fig. 5. The pristine h-BP structure exhibits the highest tensile strength, reaching approximately 65 GPa at a strain of 26.28% in Zigzag direction loading. The stress increases continuously until the peak is reached, followed by a sharp decrease associated with rapid fracture. This response indicates that the uninterrupted hexagonal B–P network can effectively distribute the applied tensile load and delay crack initiation. The introduction of an SW defect reduces the maximum tensile stress to approximately 59.34 GPa at a strain of 21.1%. Relative to pristine h-BP, this corresponds to reductions of approximately 10.3% in tensile strength and 23.8% in failure strain. Although the SW defect does not create an open vacancy, the local rearrangement of the atomic structure distorts the surrounding B–P bonds and creates a preferred location for stress concentration. The atomistic configurations in Fig. 5c–5d show that fracture develops from the SW-defected region at a strain of approximately 21.4%. The decrease in failure strain is therefore more pronounced than the decrease in strength, indicating that the SW defect has a particularly strong influence on the deformation behavior of h-BP.

A single-atom vacancy causes a degradation in mechanical performance. The vacancy-containing structure reaches a maximum tensile stress of approximately 54.91 GPa at a strain of 18.6%, representing reductions of approximately 17.0% in tensile strength and 32.9% in failure strain compared with the pristine structure. Removal of an atom interrupts the continuous B–P bonding network and reduces the number of bonds available to transfer the applied load. Consequently, stress becomes concentrated around the vacancy boundary, leading to localized bond rupture and crack initiation. As illustrated in Fig. 5e–5f, the crack originates near the single vacancy and expands through the surrounding lattice at approximately 18.0% strain. The two-atom vacancy produces the most severe degradation among the investigated defect configurations. Its maximum tensile strength decreases to approximately 50.67 GPa at a strain of 16.4%. These values correspond to reductions of approximately 23.4% in tensile strength and 40.8% in failure strain relative to pristine h-BP. Compared with the single-atom vacancy, the two-atom vacancy causes an additional reduction of approximately 7.7% in strength and 11.8% in failure strain. The larger defective region contains fewer load-bearing bonds and produces a stronger local stress concentration. The atomistic configuration in Fig. 5g–5h confirms that crack growth develops from the two-vacancy region at approximately 16.4% strain.

The abrupt stress reduction following the maximum stress in all four curves indicates the presence of a defect changes the location and strain at which fracture initiates. In the pristine structure, crack formation requires relatively high deformation because no pre-existing weak region is present. In contrast, the SW and vacancy defects act as crack-nucleation sites, causing premature failure. The degradation follows the order pristine < SW defect < single vacancy < two-atom vacancy, with the two-atom vacancy producing the lowest tensile strength and failure strain. Overall, these findings demonstrate that defects have a considerably greater effect on tensile strength and fracture strain than on the initial elastic response of h-BP. Therefore, minimizing vacancy formation will be particularly important for maintaining the structural integrity and mechanical reliability of h-BP-based devices.

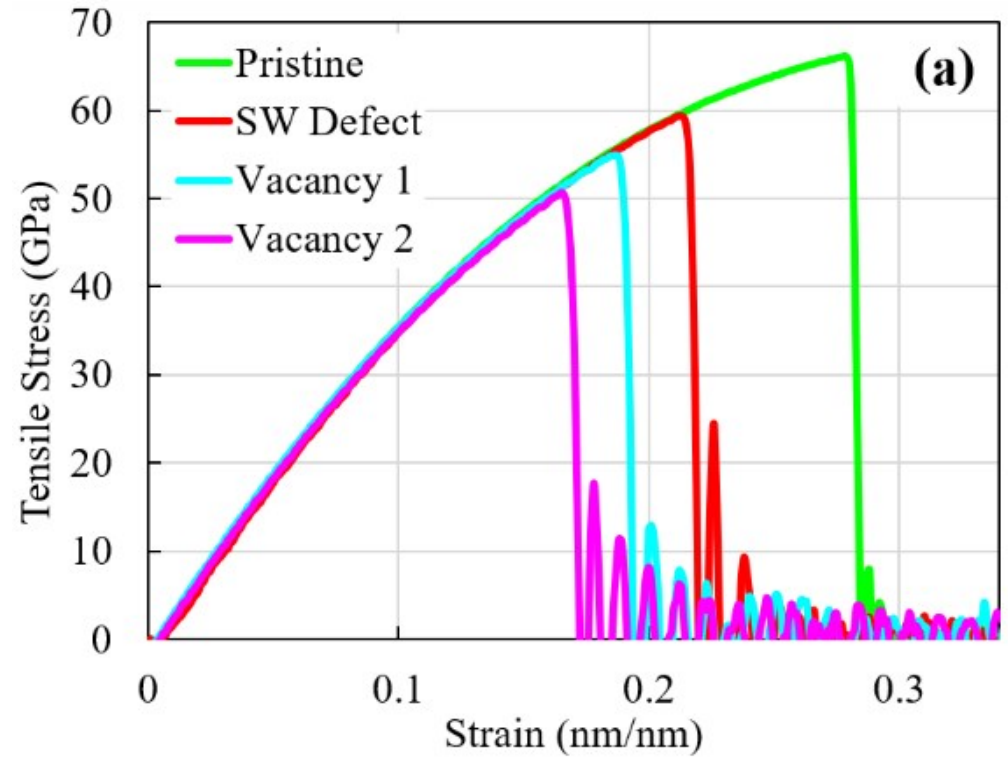

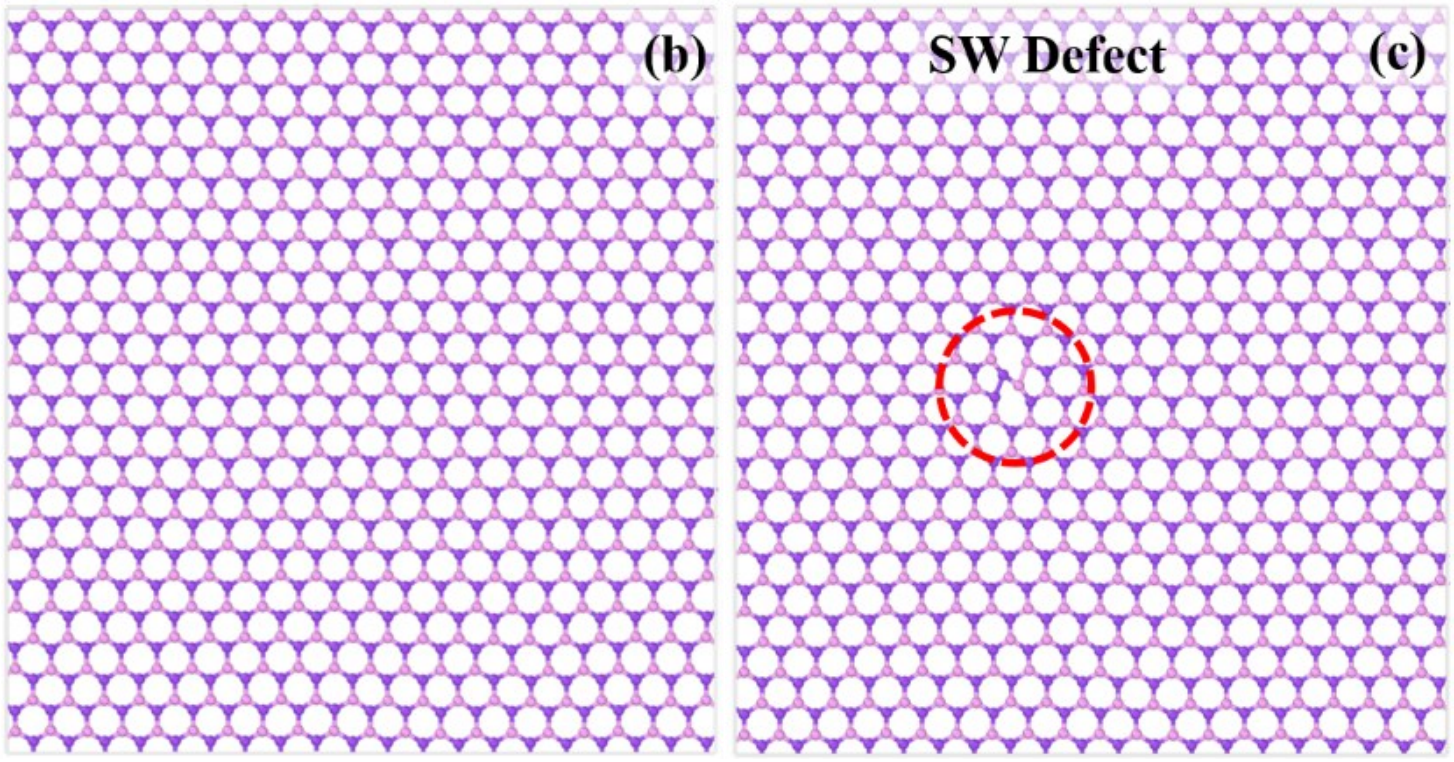

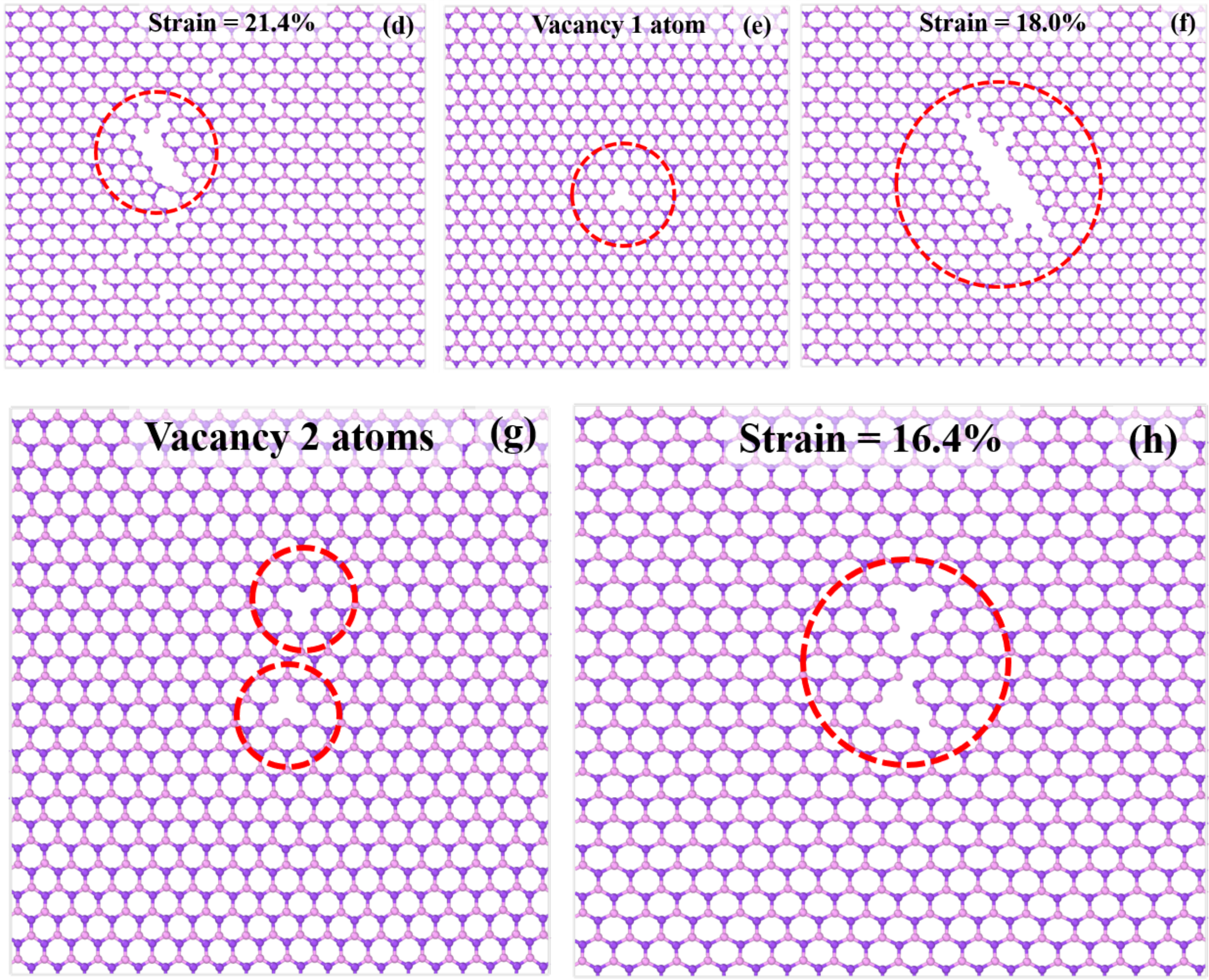

**Figure 5.** Effect of structural defects on the tensile behavior of monolayer h-BP: (a) stress–strain responses of pristine, Stone–Wales-defected, single-vacancy, and two-vacancy structures; (b) pristine h-BP; (c) and (d) Stone–Wales defect and corresponding fracture evolution; € and (f) single-atom vacancy and fracture evolution; and (g) and (h) two-atom vacancy and fracture evolution.

**4.0 Conclusion**

In this study, the thermo-mechanical properties of h-BP were investigated using MD simulations based on forcefield parameters derived from DFT calculations. The key findings of the study are summarized as follows:

**i. Mechanical anisotropy:** h-BP exhibits anisotropic mechanical behavior under tensile loading, with the zigzag direction consistently demonstrating higher tensile strength and elastic modulus compared to the armchair direction. This difference is attributed to the atomic arrangement and bonding characteristics in the two directions.

**ii. Temperature-dependent properties:** Both tensile strength and elastic modulus decrease with increasing temperature, indicating thermal softening. The tensile strength in the zigzag direction reduces from approximately 65 GPa at 300 K to around 60 GPa at 900 K, while in the armchair direction, it declines from 54 GPa to about 49.5 GPa over the same temperature range.

**iii. Elastic modulus variation:** The elastic modulus shows a similar temperature-dependent reduction, with the zigzag direction retaining higher stiffness than the armchair direction across all tested temperatures. The degradation rates indicate greater stiffness loss in the armchair direction under thermal loading.

**iv. Fracture behavior:** At room temperature, h-BP demonstrates a sudden failure after the elastic limit. As temperature increases, atomic vibrations promote earlier bond rupture, leading to reduced tensile strength and strain-to-failure.

**v. Effect of structural defects:** Structural defects significantly reduce the tensile strength and failure strain of h-BP by introducing localized stress concentrations and promoting premature crack initiation. Compared with pristine h-BP, the Stone–Wales defect reduces tensile strength and failure strain by approximately 10.3% and 23.8%, respectively, while a single-atom vacancy results in reductions of approximately 17.0% and 32.9%. The two-atom vacancy produces the most severe degradation, decreasing tensile strength by approximately 23.4% and failure strain by 40.8%. Crack propagation preferentially initiates from the defective regions, demonstrating that increasing vacancy size progressively compromises the mechanical integrity of h-BP.

These findings provide valuable insights into the structural integrity and thermal stability of h-BP. This research lays the groundwork for future experimental studies and practical applications of h-BP in electronics, energy systems, and sensing technologies.

## 5.0 Acknowledgement

ZI is thankful to the National Science Foundation (NSF) Award Number: 2206952. ZI is also thankful to Bowling Green State University, School of Engineering Faculty Start-Up Grant 33900032: SU Islam and NOIC grant 11260177GR.

## 6.0 Competing interests

On behalf of all authors, the corresponding author states that there is no conflict of interest.

## 7.0 Data Availability

The datasets used and/or analyzed during the current study available from the corresponding author on reasonable request.